\documentclass[12pt]{iopart}
\begin{document}

\title[The world problem]{The world problem:  
on the computability of the topology of 4-manifolds}

\author{J R van Meter}
\address{ NASA Goddard Space Flight Center, Greenbelt, MD  20771}

\begin{abstract}
Topological classification of the 4-manifolds bridges computation theory and physics.  A proof of the undecidability of the
homeomorphy problem for 4-manifolds is outlined here in a clarifying way.  It is shown that an arbitrary Turing machine with
an arbitrary input can be encoded into the topology of a 4-manifold, such that the 4-manifold is homeomorphic to a certain
other 4-manifold if and only if the corresponding Turing machine halts on the associated input.  Physical implications are
briefly discussed.
\end{abstract}
\pacs{0210-v, 0240Re, 0420Gz} 
\maketitle

\section{Introduction}
\label{Introduction}

A theorem proved by Markov on the non-classifiability of the 4-manifolds implies that, given some comprehensive
specification for the topology of a manifold (such as its triangulation, a la Regge calculus, or instructions for constructing it via cutting and
gluing simpler spaces) there exists no general algorithm to decide whether the manifold is homeomorphic to some other
manifold \cite{markov60}.  The impossibility of classifying the 4-manifolds is a well-known topological result, the proof of which,
however, may not be well known in the physics community.  It is potentially a result of profound physical implications, as
the universe certainly appears to be a manifold of at least four dimensions.  The burgeoning quest for the topology of the
universe \cite{luminet03} is still in its infancy; Markov's theorem may ultimately bear upon what can be deduced about it.  Already Markov's theorem
impacts certain approaches to quantum gravity.  On the basis of this theorem, and consideration of hypothetical quantum superpositions of manifolds, Penrose has 
heuristically argued that the universe
is fundamentally non-computable \cite{penrose94}.  As another example, in analogy with Feynman's sum over histories approach to quantum mechanics,
the Euclidean path integral approach to quantum gravity requires a sum over all possible topologies, with appropriate weighting, in order to calculate expectation values.  However, Markov's theorem implies inherent difficulties in
computing such a summation, as it would be impossible to decide whether a particular topology had been counted more than
once \cite{geroch86}.  

Owing to its theorized physical significance, the computability and tractability of this sum over topologies has received some attention in the literature.  Although direct summation of the series is non-computable, it is unknown whether it might
nonetheless be deducible by indirect
means, perhaps as the computable limit of some sequence; failing in that, it has been implied that the sum can nevertheless be approximated to any desired order
of accuracy \cite{geroch86}.  However, without a systematic way to proceed, there is no guarantee that such an approximation could be carried through in finite time.
To obviate such difficulty, it has been proposed to relax the condition of homeomorphy, when classifying the manifolds, and instead classify them
according to
a weaker condition, in terms of their triangulation \cite{schleich93a,schleich93b}.  But such a classification scheme would keep infinite redundancy of physically distinct
manifolds
in the series and it is not clear how to interpret the resulting sum.  More recently, partly sidestepping the issue of computability,
deductions have been made about the density of topologies per ``normalized volume" -- a geometric quantity -- in the context of a saddle-point approximation to the Euclidean path integral \cite{carlip97,carlip98,anderson04}.  The above work was motivated by 
the tantalizing possibility that this sum over topologies might determine the value of the cosmological constant
\cite{carlip97,carlip98,anderson04,coleman88}.

Manifold non-classifiability represents a fascinating juxtaposition of theoretical computer science with physics.  The
intent here is to outline a proof that will establish a correspondence between Turing machines and 4-manifolds such that
deciding whether a manifold is homeomorphic to a certain other manifold is tantamount to deciding whether the corresponding
Turing machine halts; to the author's knowledge this illuminating point has not been explicitly made elsewhere.  It is 
further hoped that the proof sketched here will provide insight into the physical relevance of Markov's theorem.

This paper is organized as follows.  In Section~2, Turing machines, and the unsolvability of the halting problem, are
reviewed.  In Section~3 it is shown that if the group triviality problem could be solved then the halting problem could be
solved.  In Section~4 it is shown that if the 4-manifold homeomorphy problem could be solved then the group triviality
problem could be solved.  These results are
discussed in Section~5.

\section{Turing machines}
\label{Turing}

A Turing machine is a formal idealization of a computer \cite{turing36}.  In its simplest formulation, a Turing machine consists of a
linear tape divided into squares onto which symbols have been printed, and a movable head that scans each square one at a
time. The sequence of symbols initially printed on the tape can be considered the input of
the Turing machine. The head can overwrite the current scanned square, move one square to the right, or move one square to the left,
depending on its internal state and its programmed instructions.  Let the $h+1$  possible states of the machine be denoted by
$q_0,...q_h$ and the $k+1$  possible symbols printed on the tape be denoted by $s_0,...s_k$.  The instructions followed by the machine 
can be conceived as a list
of if-then statements of the form: ``if the current state is $q_i$  and the current scanned symbol is $s_j$  then [either move a square
or print a symbol] and change to state $q_k$ ".  After updating its state and its current scanned symbol, the machine repeats the
process, reviewing the list of if-then statements. This goes on forever or until the machine arrives at a $(q_i,s_j)$  pair for which
it has no instructions, at which point it halts.  Note that, although more properly referred to as a program, by convention
the term ``Turing machine" is taken to be synonymous with its hardwired instructions.  

Consider, as Turing did, machines designed to output a sequence of symbols, potentially never ending, as the digits of a
real number.  Its output can be printed on every other square of the tape, while the rest of the squares are reserved for
``scratch paper".  Rather than print the entire sequence continuously, these machines will print only $j$ digits, given the
integer $j$ as an input (i.e., initially printed on some of the tape squares).  All such machines, which input an integer and output a digit, can themselves be ordered and
numbered by integers.  Turing provided a specific way to encode the instructions which
uniquely characterize each Turing machine into the digits of a (very large) integer; these integers
can then be ordered and renumbered by consecutive integers -- call them $\tau_i$.  

A Turing machine, which can examine another Turing machine by reviewing the latter's specifications on tape, cannot in
general decide whether an arbitrary Turing machine will complete its computation and halt on a given input, or go into an
infinite loop without ever printing any output.  This assertion can be proven by contradiction.  Assume the existence of a 
machine algorithm that decides, in a finite
number of steps, whether a given machine will halt on a given input.  A machine $\delta$  can then be constructed which, given an
input integer $n$, operates as follows.  $\delta$  initializes a counter $j$ to 1, checks to see whether $\tau_1$  halts on input 1, and if so
increments $j$ by 1.  $\delta$  then checks to see whether each subsequent machine $\tau_i$  halts on input $j$, in order, incrementing $j$ for
each halting machine.  When $\tau_i$ is determined not to halt, $j$ remains at the same value and the next machine $\tau_{i+1}$ is checked.  Finally, $\delta$  checks to see whether $\tau_n$  halts on input $j$, where $j$ now equals one plus the number of
halting machines up through $\tau_{n-1}$.  If $\delta$  decides that $\tau_n$  halts, then $\delta$  prints the $j$th digit computed by $\tau_n$  and then halts itself.
Otherwise, $\delta$ just halts.
Note that, in the former case, as part of $\delta$'s assigned task, $\delta$ must effectively emulate machine $\tau_n$.  (Turing proved it is possible to design a machine such as
$\delta$ to emulate any other arbitrary machine $\tau_i$ on command.)
By assumption, $\delta$ can perform all of the above operations in a finite number of steps.  

Since $\delta$ is essentially a machine that outputs a digit on being input an integer, $\delta$ itself ranks among the $\tau_i$ machines described previously.
Now give $\delta$
input $k$, such that the $k$th halting machine is $\delta$.  $\delta$ will proceed by computing the first digit output from the first halting machine, the first two
digits output from the second halting machine, and so forth, up to the first $k-1$ digits output from the $(k-1)$th halting
machine.  In so doing, $\delta$  will have computed the first $k-1$  digits of its own output sequence.  Now $\delta$ must compute the first $k$
digits of the $k$th halting machine, itself.  According to the algorithm by which $\delta$ is defined, $\delta$ must recompute the first $k-1$
digits of its output sequence.  Then to compute the $k$th digit, $\delta$  must recompute the first $k-1$  digits of its output
sequence.  And so forth, ad infinitum.  We have arrived at a contradiction:  the assumption that $\delta$ will halt on all input implies that $\delta$  will not halt on at least one input.  

Alternatively, the unsolvability of the halting problem can be understood using Cantor's diagonal argument.  If one
attempts to enumerate all of the sequences computed by halting machines, i.e. put them on a one-to-one correspondence with
the integers, one can always use a machine such as $\delta$  to construct a sequence not on the list - i.e., $1-\delta(j)$, if the output
digits are binary digits.  This would imply that the computable sequences are uncountably infinite and, as there is at
least one Turing machine for each such sequence, that Turing machines are also uncountable.  However, since Turing machines
are finitely specified, they must be countable: a contradiction, proving again that the halting problem is unsolvable. 
 
\section{Semigroups and groups}
\label{Semigroups}

A few definitions are in order.  A semigroup is a set of elements for which a binary operator has been defined so as to
satisfy closure and associativity; equivalently, it is a group in which elements are not required to have inverses.  A finitely generated semigroup or
group, generally infinite, albeit discrete, and non-Abelian, has a finite alphabet of generators.  Its elements can be
represented as ``words", i.e. strings ``spelled out" by products of generators.  A finitely presented semigroup or group is
specified by a finite number of generators and a finite number of relations, where relations are equations between words.
The word problem for semigroups or groups is the problem of finding a general algorithm which, by successive application of
the relations, can decide whether two arbitrary words are equal (in a finite number of steps).  
 
The following proof of the unsolvability of the semigroup word problem proceeds very much like that of Post \cite{post47} but has been
modified to connect it more directly with the halting problem.  Consider a semigroup $\Gamma_\tau$  with generators $q_0,q_1,...q_h$,
 $s_0,s_1,...s_k$, and $l$.  Each $q_i$
will represent a state of a Turing machine, each $s_j$  will represent a symbol on the tape, $s_0$  will represent a blank, and $l$ will
represent the left and right bounds of the string of symbols input to the machine.  

All of the operations of a Turing
machine $\tau$  can then be represented by relations in $\Gamma_\tau$.  The action of printing over symbol $s_b$ with symbol $s_d$ can be represented by the
following relation,
\begin{equation}
\label{print}
q_a s_b  = q_c s_d 
\end{equation}
where $a$ and $c$ have some specific values between 1 and $h$, and likewise $b$ and $d$ between 0  and $k$.  In accordance with Turing's 
convention, all machine actions will be accompanied by a simultaneous change of state.  Similarly, the action of moving to the
left one space can be represented by the following $h+2$  relations.
\begin{eqnarray}
s_iq_as_b = q_cs_is_b,\quad i=0,1,...,h \\ 
lq_as_b = lq_cs_0s_b 
\end{eqnarray}
And the action of moving to the right one space can be represented by the following $h+2$ relations.
\begin{eqnarray}
q_as_bs_i = s_bq_cs_i,\quad i=0,1,...,h \label{right_a} \\ 
q_as_bl = s_bq_cs_0l \label{right_b} 
\end{eqnarray}
This completes the semigroup ``emulation" of a Turing machine.  

For the purpose of investigating the halting problem, I'm going to introduce two new generators with the unconventional
notation $\rangle$  and $\langle$ , for reasons that will soon become clear.  For every $q_as_b$  pair that does not appear in the left hand side of
equations (\ref{print}-\ref{right_b}), add the relation:
\begin{equation}
q_as_b=\rangle s_b
\end{equation}
Now add the following $2h+3$ relations:
\begin{eqnarray}
s_i \rangle=\rangle,\quad i=0,1,...,h \label{rangler} \\
l\rangle = l\langle \\
\langle s_i = \langle, \quad i=0,1,...,h \label{langler}
\end{eqnarray}
In effect,  $\rangle$  devours all symbols to its left.  If it comes to the end-marker $l$, it mutates into $\langle$ .  $\langle$  devours all symbols
to its right.  

The outcome is that if any word $\omega_\iota$  corresponds to an input $\iota$  on which the associated Turing machine halts, then it can be
shown to be equivalent, by repeated application of the above relations (\ref{rangler}-\ref{langler}), to the word $l\langle l$.  If a word does not
correspond to an input on which the associated Turing machine halts, then it is not equivalent to the word  $l \langle l$.   By
convention, $q_0$  is reserved for the halting state, so the relation $l \langle l = q_0$  might be added - then $\omega_\iota = q_0$  in 
$\Gamma_\tau$  if and only if $\tau$  halts on
input $\iota$.   An algorithm that could solve the word problem for semigroups, therefore, could solve the halting problem for
Turing machines.

The above result for semigroups has direct implications for groups.  For each finitely presented semigroup $\Gamma_\tau$ described
above, there is a prescription for constructing a finitely presented group $G'_\tau$  such that for every generator and
relation in $\Gamma_\tau$  there is a corresponding generator and relation in $G'_\tau$, and the following theorem holds:  There exist words $u_\iota$ and
$v_\iota$ in the finitely presented group $G'_\tau$ that are equal if and only if $\omega_\iota=q_0$  in the finitely presented semigroup $\Gamma_\tau$ \cite{boone59}.  Equivalently,
$w_\iota \equiv u_\iota v_\iota^{-1}=1$ in $G'_\tau$  if and only if $\omega_\iota = q_0$  in $\Gamma_\tau$.  Further, for each finitely 
presented group $G'_\tau$  and each word $w_\iota$  in $G'_\tau$  there is a prescription for
constructing a finitely presented group $G_\tau(w_\iota)$  such that for every generator and relation in $G'_\tau$  there is a corresponding
generator and relation in $G_\tau(w_\iota)$  and the following theorem holds:  $G_\tau(w_\iota)$  is trivial, i.e. contains only the identity element, if and only if $w_\iota = 1$ in $G'_\tau$ \cite{nabutovsky96}.  It follows that the
triviality of finitely presented groups is algorithmically undecidable. 

\section{Manifolds}
\label{Manifolds}

Each element of the fundamental group of a manifold represents an equivalence class of closed paths in the manifold that can be
continuously deformed into one another, i.e., a homotopy class of closed paths.  As an example, a trivial element in a fundamental group represents a class of paths
that can be contracted to a point, and a trivial fundamental group implies a simply connected manifold.  As
another example, the infinite cyclic group, which can be finitely presented by one generator and no relations, is the
fundamental group of a hypersphere with one arcwise connected handle: each element of the group, equal to the generator
raised to some power $p$, corresponds with the homotopy class of paths that wind about the handle $p$ times (and negative
powers will be said to correspond to counterwindings, described below).  It will be shown that for any given finitely presented group, a
manifold can always be constructed for which the given group is fundamental.  The prescription can be summarized as
attaching to a hypersphere a handle for each generator of the group, followed by further surgery to accommodate each
relation.  

The following construction is homeomorphic to that of Markov, but the method of construction has been streamlined
for pedagogical purposes.  
Consider an arbitrary finitely presented group of the form 
\begin{equation}
G=\{g_1,...,g_m|r_1,...,r_n\}
\end{equation}
where each $r_i$  is a word representing a relation of the form
$r_i=1$ and is called a relator.  Beginning with the 4-sphere, $S^4$, for each generator $g_i$  attach a handle of the form $H_i=S^3 \times [-1,+1]$.  
Each such attachment is performed by removing from $S^4$ two non-intersecting, open 4-balls and identifying the resulting 3-spherical boundaries with the ends
of $H_i$.
Calling the
former $S^4$  region $A$, the attachments are subject to the conditions that no two handles intersect, and the intersection of
each handle with $A$ is a union of two 3-spheres:  $H_i \cap H_j=0$, $i \neq j$, $A \cap H_i=S^3 \times \{-1,+1\}$.  
In this manner a manifold can be handily constructed for each
free fundamental group of the form $\{g_1,...,g_n|\}$.  To understand this, note that the construction thus far is homeomorphic to the
connected sum of $m$ copies of $S^3 \times S^1$, then use the fact that the fundamental group of the cross product of manifolds is the free
product of the fundamental groups of the manifolds, while the fundamental group of the connected sum of manifolds is the
direct product of the fundamental groups of the manifolds.

 An arbitrary word can be represented by a closed path in the above construction as follows.  Consider a path that begins at
some point inside $A$.  Reading the word from left to right, represent each generator $g_i$  of positive power $p$ by a path that
enters its associated handle $H_i$  at $S^2 \times \{-1\}$ , then exits $H_i$  at $S^2 \times \{+1\}$ , then circles back around and repeats $p-1$ times.  Represent negative
powers $-p$ the same way but switch $S^2 \times \{-1\}$  and $S^2 \times \{+1\}$  (hence negative powers ``unwind" positive powers).  After exiting the handle for
the $p$th time, continue the path to the handle associated with the next generator in the word, and repeat the winding
process, continuing in this way until the last generator in the word has been represented.  Finally, join the end of the
path with its starting point to close the loop.  

A relator of the finitely presented fundamental group, being a word equated with the identity, corresponds to paths that can
be continuously deformed to a point.  Obviously such deformation of a path through a handle is obstructed; some topological
surgery will be necessary to bypass the obstruction.  For each relator $r_j$ , gouge out a region from the above constructed
manifold (call the manifold $M$) along the vicinity of a path representative of $r_j$  such that the gouged-out region is
homeomorphic to $U^3 \times S^2$ , where $U^3$ is the open 3-ball.  Simultaneously, in a copy of $S^4$ , gouge out a similar $U^3 \times S^1$  region; call this
manifold  $O_j$.  Finally, identify the $S^2 \times S^1$  boundary of the gouged-out region in $M$ (call this boundary $T_j$ ) with the $S^2 \times S^1$  boundary of
the gouged-out region in $O_j$.  Note that $O_j$  is simply connected.  (To see this, consider that the
only conceivably non-trivial closed path in $O_j$ is one that interlocks with the loop formed by the gouged-out region.  But
the former can be continuously deformed to the boundary of the latter, whereupon it can be made to encircle a
cross-section homeomorphic to $S^2$, and thereon contracted to a point.)  Any path in the homotopy class of paths associated with
the relator $r_j$  can now be continuously deformed to the surface of $T_j$ , then contracted to a point in $O_j$ .  Repeat this
surgery for each relator, in this way gluing to $M$, $n$ copies of $O_j$.  This completes the construction.  It can be verified, by
considering the fundamental groups of the subspaces that cover $M$ \cite{seifert80}, that the fundamental group of $M$ is the given group $G$  as
advertised.  

If two manifolds are homeomorphic, their fundamental groups are isomorphic.  But the converse is not necessarily true, thus
the non-classifiability of the manifolds does not immediately follow from the non-classifiability of their fundamental
groups.  Fortunately for the purposes of this proof, the manifolds constructed above have the following critical property.
First consider another manifold formed by gouging out from $S^4$, $m$ non-intersecting
regions homeomorphic to $U^3\times S^1$, and gluing the remaining boundaries to those of an
identical copy; call the
resulting manifold $N_m$.   Given one of the previously constructed manifolds $M$ such that its fundamental group $G$ has $m$
generators, if $G$ is trivial then, it turns out,  $M$ must be homeomorphic to $N_m$ \cite{markov60}.  

To come full circle, let the fundamental group of the manifold $M$ represent a Turing machine:  let $M=M_\tau(w_\iota)$  such that its
fundamental group is $G_\tau(w_\iota)$ , as described in Section~2.  Call $M_\tau(w_\iota)$  a Turing manifold.  Call $N_{m(\tau,\iota)}$, where $m(\tau,\iota)$ is the number of generators
required to represent the Turing machine $\tau$  with input $\iota$  by $G_\tau(w_\iota)$, a halting manifold.  It follows that the Turing manifold $M_\tau(w_\iota)$  is
homeomorphic to the halting manifold $N_{m(\tau,\iota)}$  if and only if Turing machine $\tau$  halts on input $\iota$ .  

\section{Discussion}
\label{Discussion}

A sketch of a proof has been given for the non-classifiability of the 4-manifolds, by way of a topological construction
whereby a 4-manifold represents a Turing machine.  More precisely, a Turing machine has been encoded into a finitely
presented semigroup, which has been encoded into a finitely presented group, which along with a particular Turing input has
been encoded into another finitely presented group, which has been encoded into a 4-manifold. The chain of encodings is such
that solving the homeomorphy problem for 4-manifolds would solve the halting problem for Turing machines, which is
unsolvable.  
Expressed more intuitively, the essence of the problem is that the topology of a 4-manifold is potentially so rich that its
complexity can rival that of any computer program intended to analyze it.  Inputting the specifications of a 4-manifold to 
such a computer program can, in a sense, be equivocated with inputting a computer program to a computer program --
an enterprise subject to logical paradoxes and limitations of the kind brought to light by Turing.

Regarding the physical applicability of Markov's theorem, while the constructions considered above are compact 4-manifolds,
spacetime is often considered to be non-compact, and is sometimes speculated to have hidden extra dimensions.  Markov's
proof applies equally well to higher dimensional manifolds - consider $M \times S^{d-4}$, where $d>4$  - as well as non-compact manifolds -
consider $M\#R^4$.  Granted Markov's theorem only applies to manifolds that are permitted to be non-simply connected, but there is
a strong possibility that the universe lives in this category.  On the cosmic scale, the universe may be
multiply-connected \cite{luminet03}; on the stellar scale, black hole interiors may be topologically nontrivial, though such
nontriviality might be rendered undetectable by event horizons \cite{friedman93} (on the other hand, traversable worm holes might exist \cite{morris88}); on the subatomic scale, particles
are sometimes speculated to be topological geons \cite{diemer99,hadley00};
and on the Planck scale, spacetime foam is conjectured to perturb the local topology to no end \cite{carlip97,carlip98,anderson04,wheeler57}.  

It is conceivable that some physical criteria could be found which would restrict permissible 4-manifolds to classifiable manifolds.
For example, if a strict interpretation of causality is imposed, in the form of the conditions of isochrony and the exclusion of closed timelike curves, then it can be shown that the allowed 
4-manifolds are constrained to those of the form 
$C\times [0,1]$, $C\times [0,\infty)$, and $C\times(-\infty,\infty)$, where $C$ is a 3-manifold \cite{geroch67}.  These manifolds are classifiable if the 3-manifolds are classifiable;
although whether the 3-manifolds are classifiable is still an open question.  
Note that the proof of Markov's theorem, as sketched above, is not applicable to 3-manifolds; for example, the three-dimensional analog of $O_j$ is not simply connected,
as required.  In a sense, there is not enough ``room" in a 3-manifold to topologically encode a Turing machine, and so there is hope that 3-manifolds might be classifiable.
However, whether the universe obeys the previously mentioned interpretation of causality is unknown.  These particular conditions may be too restrictive; they would
preclude Wheeler's spacetime foam, as well as other exotic but physically motivated topological proposals.
In summary, on the basis of
current physical knowledge, the non-classifiability of the 4-manifolds remains relevant.  

\ack
I wish to thank Steve Carlip for inspiring discussion.  This work was supported in part by a National Research Council
Associateship Award at the Goddard Space Flight Center, funded by NASA Space Sciences grant ATP02-0043-0056, and in part by
Department of Energy grant DE-FG02-91ER40674.

\end{document}